# Computation of a Tree 3-Spanner on Trapezoid Graphs

*Sambhu Charan Barman[1], Sukumar Mondal[2] and Madhumangal Pal[1]*
[1]Department of Applied Mathematics with Oceanology and Computer Programming,
Vidyasagar University, Midnapore - 721 102, India.
email: {barman.sambhu, mmpalvu}@gmail.com

[2]Department of Mathematics, Raja N. L. Khan Women's College, Gope Palace,
Midnapur - 721 102, India.
email: sm5971@rediffmail.com

**Abstract.** In a graph $G$, a spanning tree $T$ is said to be a tree t-spanner of the graph $G$ if the distance between any two vertices in $T$ is at most $t$ times their distance in $G$. The tree t-spanner has many applications in networks and distributed environments. In this paper, an algorithm is presented to find a tree $3$-spanner on trapezoid graphs in $O(n^2)$ time, where $n$ is the number of vertices of the graph.

## 1. Introduction
### 1.1. Trapezoid graph
A *trapezoid graph* can be represented in terms of *trapezoid diagram*. A *trapezoid diagram* consist of two horizontal parallel lines, named as top line and bottom line. Each line contains $n$ intervals. Left end point and right end point of an interval $i$ are $a_i$ and $b_i (\geq a_i)$ on the top line and $c_i$ and $d_i (\geq c_i)$ on the bottom line. A *trapezoid i* is defined by four corner points $[a_i, b_i, c_i, d_i]$ in the trapezoid diagram. Let $T = \{1, 2, \ldots, n\}$, be the set of $n$ trapezoids. Let $G = (V, E)$ be an undirected graph with $n$ vertices and $m$ edges and let $V = \{1, 2, \ldots, n\}$. $G$ is said to be a *trapezoid graph* if it can be represented by a trapezoid diagram such that each trapezoid corresponds to a vertex in $V$ and $(i, j) \in E$ if and only if the trapezoids $i$ and $j$ intersect in the trapezoid diagram [9]. Two trapezoids $i$ and $j(>i)$ intersect if and only if either $(a_j - b_i) < 0$ or $(c_j - d_i) < 0$ or both. We assume that the graph $G = (V, E)$ is connected. Without any loss of generality we assume the following :
($a$) a trapezoid contains four different corner points and that no two trapezoids
 share a common end point,
($b$) trapezoids in the trapezoid diagram and vertices in the trapezoid graph are one and same thing,
($c$) the trapezoids in the trapezoid diagram $T$ are indexed by increasing right end points on the top line i.e., if $b_1 < b_2 < \cdots < b_n$ then the trapezoids are indexed by $1, 2, 3, \cdots, n$ respectively.

      Figure 2 represents a trapezoid graph and it's trapezoid representation is

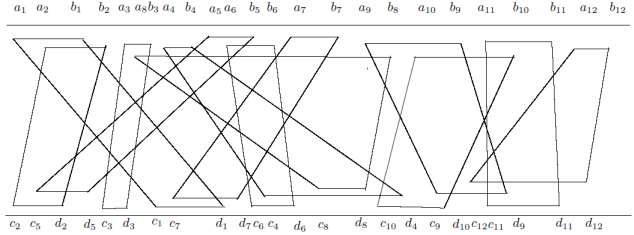

**Figure 1:** A trapezoid diagram T of the graph G of Figure 2.

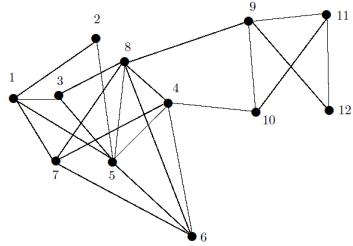

**Figure 2:** A trapezoid graph G.

shown in Figure 1. The class of trapezoid graphs includes two well known classes of intersection graphs: the permutation graphs and the interval graphs [11]. The permutation graphs are obtained in the case where $a_i = b_i$ and $c_i = d_i$ for all $i$ and the interval graphs are obtained in the case where $a_i = c_i$ and $b_i = d_i$ for all $i$. Trapezoid graphs can be recognized in $O(n^2)$ time [13]. The trapezoid graphs were first studied in [8, 9]. These graphs are superclass of interval graphs, permutation graphs and subclass of cocomparability graphs [12].

Lot of works have been done to solve different problems on graph theory, particularly on interval, circular-arc, permutation, trapezoidal, etc. graphs [22-41].

### 1.2. Definitions

Let $G = (V, E)$ be a graph with vertex set $V$ and edge set $E$, where $n$ be the number of vertices in $V$ and $m$ be the number of edges in $E$. The *distance* between two vertices $u$ and $v$ in $G$ is denoted by $d_G(u,v)$ and it is the minimum number of edges required to traversed from $u$ to $v$ or $v$ to $u$.

For a connected graph $G = (V, E)$, $H = (V, E')$ is a spanning subgraph iff $E' \subseteq E$. A $t$-spanner of a graph $G$ is a spanning subgraph $H(G)$ in which the distance between every pair of vertices is at most $t$ times their distance in $G$, i.e., $d_H(u,v) \leq t d_G(u,v)$, for all $u, v \in V$. The parameter $t$ is called the stretch factor. The minimum $t$-spanner problem is to find a $t$-spanner $H$ with the fewest possible edges for fixed $t$. The spanning subgraph $H$ is called a minimum $t$-spanner of $G$ and it is denoted by $H_t(G)$. A spanning tree of a connected graph $G$ is an acyclic connected spanning subgraph of $G$. A tree spanner of a graph is a spanning tree that approximates the distance between the vertices in the original graph. In particular, a spanning tree $T$ is said to be a tree $t$-spanner of a graph $G$ if the distance between every pair of vertices in $T$ is at most $t$ times their distance in $G$, i.e., $d_T(u,v) \leq t d_G(u,v)$, for all $u,v \in V$.

### 1.3. The $t$-spanner problem
The minimum $t$-spanner problem is of two types: decision version and optimization version.
The decision version of the problem is stated as follows.

**Decision Version:**
**Input**: A graph $G = (V,E)$ and $k \geq 0$ are given.
**Question**: Whether $G$ has a $t$-spanner with $k$ or fewer edges, i.e.,
$$|E(H_t(G))| \leq k.$$

The optimization version of the problem is stated as follows.
**Optimization Version:**
**Input**: A graph $G = (V,E)$.
**Problem**: Find a $t$-spanner with fewest possible edges for a fixed $t$.
In this paper, the optimization version of the problem is considered.

### 1.4. Applications of $t$-spanners
The $t$-spanner and tree $t$-spanner have many applications in communication networks, distributed systems, etc. The notion of $t$-spanner was introduced by Peleg and Ullman [17] in connection with the design of synchronizers. The synchronizer is a simulation technology introduced by Awerbuch [1] and it is used to transform synchronous algorithms into efficient asynchronous algorithms to execute on asynchronous network. The $t$-spanner is the underlying graph structure of the synchronizer, and the stretch factor and the size of the $t$-spanner are closely related to the time and communication complexities of the synchronizer respectively. Spanners also have application in planning efficient routing schemes to maintain succinct routing tables [18]. Spanners also arise in computational geometry in the study of approximation of complete Euclidean graphs [7]. In addition to this, it is used in computational biology in the process of reconstruction of phylogenetic trees [2].

### 1.5. Survey of the related works
In the construction of the spanner, the fundamental problem is to find a minimum $t$-spanner of a graph, where $t(\geq 1)$ is a fixed integer. The construction of minimum 2-spanner is NP-hard for general graphs [18]. In [4], Cai showed that the construction of $t$-spanner is NP-hard for each $t \geq 3$. Determination of minimum $t$-spanner for each fixed $t \geq 2$, is still NP-hard on graphs with maximum degree equal to 9 [5]. Madanlal et al. [14] have designed linear time algorithms to find minimum $t$-spanner on interval and permutation graphs for each fixed $t \geq 3$. Besides, when $t = 2$ the problem remains open for interval and permutation graphs. A linear time algorithm is designed to find a minimum 2-spanner on graphs with a bounded degree less than 4 [5]. This problem is NP-hard for perfect graphs even for chordal graphs when $t \geq 2$ [21]. However, the problem is polynomial solvable for interval graph when $t \geq 3$ [14, 15]. For $t = 2$, the exact complexity of the problem still remains open, but a polynomial time 2-approximation algorithm is available in [21]. For permutation graphs, the exact complexity of determining 2-spanners remains open, but, for $t \geq 3$ the problem is polynomial solvable [14]. For the split graph, the problem is NP-hard when $t = 2$ and polynomial solvable when $t \geq 3$ [21]. However, for the bipartite graphs the problem is trivially polynomial solvable for $t = 2$ and NP-hard for $t \geq 3$ [4]. In [14], Madanlal et al. have designed an $O(n+m)$ time sequential algorithm to find tree 3-spanner on interval graphs, permutation graphs and regular bipartite graphs, where $m$ and $n$ represent, respectively, the number of edges and vertices. Saha et al. [19] have designed an optimal parallel algorithm to construct a tree 3-spanner on interval graphs in $O(\log n)$ time using $O(n/\log n)$ processors on an EREW-PRAM. Recently, Barman et al. [3] have designed a linear time algorithm to construct a tree 4-spanner on trapezoid graphs in $O(n)$ time.

### 1.6. Main result
Here we consider the problem of determining the tree $3$-spanner on undirected, simple and connected trapezoid graphs. In this paper, we design an algorithm to construct a tree $3$-spanner on trapezoid graphs in $O(n^2)$ time, where $n$ is the number of vertices.

### 1.7. Organization of the paper
In the next section, i.e. in Section $2$, we shall discuss about BFS tree of trapezoid graphs and the main path between the vertices $1$ and $n$. In Section $3$, we present the algorithm of marking all alternative shortest paths between the root $1$ and the members of the last level of the BFS tree. Some notations have also presented in this section. Some important results related to tree $3$-spanner on trapezoid graphs are also investigated, in Section $4$. In section $5$, we discuss about the modified main path and the algorithm for finding tree $3$-spanner of the trapezoid graph. The time complexity is also calculated in this section.

## 2. The BFS tree and the main path
### 2.1. The BFS tree
It is well known that the BFS is an important graph traversal technique. It also constructs a BFS tree. The BFS, started with an arbitrary vertex $v$. We visit all the vertices adjacent to $v$ and then move to an adjacent vertex $w$. At $w$ we then visit all vertices adjacent to $w$ which is not visited earlier and move to an adjacent vertex of $w$. If all the vertices adjacent to $w$ are already visited then go back to the vertex $v$ and select a vertex adjacent to $v$, which is unvisited. This process is continued till all the vertices in the graph are considered [10].

     A BFS tree can be constructed on general graphs in $O(n+m)$ time, where $n$ and $m$ represent respectively the number of vertices and number of edges of the graph [20]. Recently, Mondal et al. [16] have designed an algorithm to construct a BFS tree $T^*(i)$ with root as $i \in V$ on trapezoid graph $G = (V, E)$ in $O(n)$ time, where $n$ is

Computation of a Tree 3-Spanner on Trapezoid Graphs

the number of vertices. A BFS tree $T^*(1)$ rooted at 1 of the trapezoid graph of Figure 2 is shown in Figure 3.

We define the *level* of a vertex $v$ as a distance of $v$ from the root 1 of the tree $T^*(1)$ and denoted by $level(v), v \in V$ and take the level of root 1 as 0. The level of each vertex on BFS tree $T^*(1)$, $1 \in V$ can be assigned by the BFS algorithm of Chen and Das [6].

Let $h$ be the height of the tree $T^*(1)$. The set of all vertices at level $i$ of $T^*(1)$ is denoted by $L_i$, i.e., $L_i = \{u : level(u) = i\}$.

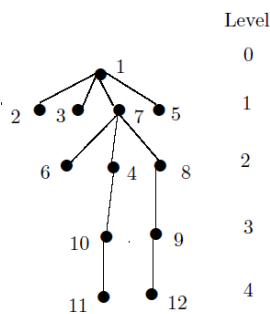

**Figure 3:** A BFS tree $T^*(1)$ of the graph G of Figure 2.

## 2.2. Computation of the main path on the BFS tree $T^*(1)$

In the BFS tree $T^*(1)$, rooted at 1, let the distance between 1 and $n$ be $k$, i.e., $level(n) = k$, where $k$ is a fixed positive integer. Also we assume that $1 \to z_1 \to z_2 \to \cdots \to z_{k-1} \to n$ be the shortest path between 1 and $n$ with 1 as parent of $z_1$, $z_i$ as parent of $z_{i+1}$ for all $i = 1,2,3,\ldots,k-2$ and $z_{k-1}$ as parent of $n$ on the BFS tree $T^*(1)$ and let this path be the *main path* between 1 and $n$.

Let $u_i'$ be the vertex on the *main path* at level $i$ on $T^*(1)$. The open neighbourhood set of any vertex $u$ is denoted by $N(u)$ and defined by $N(u) = \{x : x \in V \text{ and } (x,u) \in E\}$.

## 3. Marking of all alternative shortest paths

We mark all alternative shortest paths between the root($u_0' = 1$) of $T^*(1)$ and the members of the set $L_h$, by the following algorithm.

**Algorithm MASPT**
**Input**: The corner points $[a_i, b_i, c_i, d_i]$ of the trapezoid $i$ for all $i = 1,2,\cdots,n$.
**Output**: All marked alternative shortest paths between $u_0'$ and the members of the set $L_h$, which is a subgraph of $G = (V, E)$ and denoted by $M^*$.

**Step 1**: Compute open neighbourhood, $N(x)$, for all $x \in V$.

**Step 2**: Construct a BFS tree $T^*(1)$ of the graph $G$ with root as $1(= u_0')$.

**Step 3**: Find the sets $L_i, i = 1, 2, \cdots, h$.

**Step 4**: Mark the members of the set $L_h$.

**Step 5**: Mark all unmarked vertices at level $h-1$ which are adjacent to the marked vertices of the set $L_h$ and add the edges (if they are not present on the tree $T^*(1)$) between the marked vertices at level $h-1$ and the marked vertices at level $h$ and also mark these edges.

**Step 6**: Mark all unmarked vertices at level $h-2$ which are adjacent to the marked vertices at level $h-1$ and add the edges (if they are not connected on the tree $T^*(1)$) between the marked vertices at level $h-2$ and the marked vertices at level $h-1$ and also mark these edges and go to the next level.

**Step 7**: This process is continued until all edges between $u_0'$ and the marked vertices of level 1 are marked.

**Step 8**: Delete all unmarked vertices from BFS tree and let the reduced subgraph be $M^*$.

**end MASPT**.

The Algorithm MASPT gives the subgraph $M^*$ of $G$. A subgraph $M^*$ of the graph of Figure 2 is shown in the Figure 4. Now we calculate the time complexity of the **Algorithm MASPT**. For this purpose, we define the set $P_i$ as follows:

$P_i$: the set of marked vertices at level $i$ on $M^*$, $i = 1, 2, \cdots, h$ and let $|P_i| = l_{h-i}$ where $h$ is the height of the BFS tree $T^*(1)$).

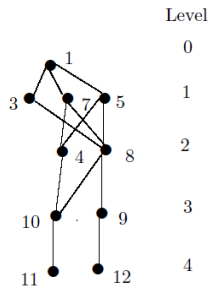

**Figure 4:** Subgraph $M^*$ of the trapezoid graph G.

**Theorem 1.** *The time complexity of marking all alternative shortest paths between the root($u_0'$) of the BFS tree $T^*(1)$ and the members of the set $L_h$, is $O(n^2)$.*

**Proof.** Step 1 can be computed in $O(n^2)$ time. In Step 2, BFS tree can be constructed in $O(n)$ time. In Step 3, computation of the sets $L_i, i = 1, 2, \cdots, h$ can be finished in $O(n)$ time. Step 4 can be completed in $O(l_0)$ time. The time complexities of Step 5, Step 6 and Step 7 are respectively $O(l_0 l_1)$, $O(l_1 l_2)$ and $O(l_2 l_3 + l_3 l_4 + \cdots + l_{h-2} l_{h-1} + l_{h-1})$. Also, Step 8 can be completed in $O(n)$ time. Hence the total time complexity of **Algorithm MASPT** is

$O(n^2) + O(n) + O(n) + O(l_0) + O(l_0 l_1) + O(l_1 l_2) +$
$O(l_2 l_3 + l_3 l_4 + \cdots + l_{h-2} l_{h-1} + l_{h-1}) + O(n)$
$= O(n^2) + O(l_0 l_1) + O(l_1 l_2 + l_2 l_3 + l_3 l_4 + \cdots + l_{h-2} l_{h-1})$
$= O(n^2) + O((1/2)(l_0 + l_1 + l_2 + \cdots + l_{h-1})^2 - (1/2)(l_0^2 + l_1^2 + l_2^2 + \cdots + l_{h-1}^2) -$
$(l_0 l_2 + l_0 l_3 \cdots + l_0 l_{h-1} + l_1 l_3 + l_1 l_4 + \cdots + l_1 l_{h-1} + l_2 l_4 + l_2 l_5 \cdots + l_2 l_{h-1} + \cdots + l_{h-3} l_{h-1}))$
$\leq O(n^2) + O((1/2)(l_0 + l_1 + l_2 + \cdots + l_{h-1})^2)$
$\leq O(n^2) + O((1/2)n^2)$ [as $l_0 + l_1 + l_2 + \cdots + l_{h-1} < n$] $\leq O(n^2)$.

Therefore, the over all time complexity of the **Algorithm MASPT** is $O(n^2)$

### 3.1. Some notations
Here we introduce some notations those are used in the rest of the paper.

h : the height of the BFS tree $T^*(1)$.

$level(v)$ : the distance of the vertex $v$ from the root 1 of $T^*(1)$, i.e.,
$d_G(1, v) = level(v)$.

$L_i$ : $L_i$ is the set of vertices at the $i$th level on the BFS tree $T^*(1)$, i.e.,
$L_i = \{x : x \text{ lies at the } i\text{th level}\}$, $i = 1, 2, \cdots, h$.

k : the length of the main path between the vertices 1 and $n$.

$u'_i$ : $u'_i$ is the vertex on the main path at level $i$.

$u^*_i$ : $u^*_i$ is the vertex on the modified main path at level $i$.

$P_i$ : $P_i$ is the set of vertices at level $i$ on the subgraph $M^*$.

$F_i$ : $F_i$ is the set of vertices which are in $L_i$ but not in $P_i$, i.e.,
$F_i = L_i - P_i$.

$S_{i,(i-1)}$ : $S_{i,(i-1)} = \{x : x \in L_i - \{u'_i\}$ and $(x, u'_i) \notin E$, $(x, u'_{i+1}) \notin E\}$

$S'_{i,(i-1)}$ : $S'_{i,(i-1)} = \{x : x \in L_i - \{u'_i\} - S_{i,(i-1)}$ and $(x, y) \in E$ where
$y \in S_{i,(i-1)}$ and $(x, u'_i) \notin E\}$.

$S''_{i,(i-1)}$ : $S''_{i,(i-1)} = \{x : x \in L_i - \{u'_i\} - S_{i,(i-1)} - S'_{i,(i-1)}$ and $(x, y) \in E$ where
$y \in S'_{i,(i-1)}$ and $(x, u'_i) \notin E\}$.

$S^*_{i,(i-1)}$ : $S^*_{i,(i-1)} = S'_{i,(i-1)} \cup S_{i,(i-1)} \cup S''_{i,(i-1)}$.

$D_i$ : $D_i = \{x : x \in S^*_{i,(i-1)}$ and $(x, y) \notin E$ where for all $y \in P_{i+1} - \{u'_{i+1}\}\}$.

$max(b_i)$ : $max(b_i) = max\{b_y : y \in P_{i+1} - \{u'_{i+1}\}, (y, u'_{i+1}) \in E$ and for all $x \in S^*_{i,(i-1)}, (x, y) \in E\}$.

$max(d_i)$ : $max(d_i) = max\{d_y : y \in P_{i+1} - \{u'_{i+1}\}, (y, u'_{i+1}) \in E$ and for all $x \in S^*_{i,(i-1)}, (x, y) \in E\}$.

$max(b^*_i)$: $max(b^*_i) = max\{b_y : y \in P_i - D_i - \{u'_i\}$ and $(x, y) \in E$ where $x \in D_i$ and $(y, z) \in E$ such that $z \in P_{i+1} - \{u'_{i+1}\}$ and $(z, u'_{i+1}) \in E\}$.

$max(d^*_i)$: $max(d^*_i) = max\{d_y : y \in P_i - D_i - \{u'_i\}$ and $(x, y) \in E$ where $x \in D_i$ and $(y, z) \in E$ such that $z \in P_{i+1} - \{u'_{i+1}\}$ and $(z, u'_{i+1}) \in E\}$.

Before going to our proposed algorithm we prove the following important results relating to tree 3-spanner on trapezoid graphs.

## 4. Some important results

In this section, according to our observations, we present some important results relating to the tree 3-spanner on trapezoid graphs.

**Lemma 1.** *The members of the set $F_i$ at any level $i$, are not adjacent with the members of the set $P_{i+1}$.*

**Proof.** Let us assume that the members of the set $F_i$ are adjacent with the members of the set $P_{i+1}$. Also we assume that $y$ be any member of the set $F_i$ and $z$ be any member of the set $P_{i+1}$. So, $(y, z) \in E$ and there is at least one path between the root $1(= u'_0)$ of the tree $T^*(1)$ and $z$ such as $z \to y \to parent(y) \to parent(parent(y)) \to \cdots \to u'_0$. This implies that $y \in P_i$ But it is impossible. Therefore the members of the set $F_i$ at any level $i$, are not adjacent with the members of the set $P_{i+1}$.

Next we consider few important results, proved by Barman et al. [3] on the BFS tree of the trapezoid graph.

**Lemma 2.**
(a) *If $i$ and $j$ are two internal nodes of same level on the BFS tree $T^*(1)$ and $b_j < b_i$ then $d_i < d_j$.*
(b) *There exists at most two internal nodes at any level on the BFS tree $T^*(1)$.*
(c) *If $i$ and $j$ are two internal nodes at any level $l$ on the BFS tree $T^*(1)$ then $(i, j) \in E$.*
(d) *If $parent(m) = j$ and $parent(k) = i$ where $i$, $j$ are two internal nodes at any*

level $l$ and $m,k$ are two vertices at level $l+1$ and also $k$ is an internal node at level $l+1$ on the BFS tree $T^*(1)$, then either $(m,k) \in E$ or $(m,i) \in E$ or both.

(e) If $parent(n) = j$ and $parent(k) = i$ where $i$, $j$ are two internal nodes at any level $l$ and $n$ (highest numbered vertex), $k$ are two vertices at level $l+1$ on the BFS tree $T^*(1)$ then either $(k,n) \in E$ or $(k,j) \in E$ or both.

(f) If $n$ be the vertex at level $l$ and $j$ be the vertex at level $l+1$ on the BFS tree $T^*(1)$, then $parent(j) = n$.

Other important results are presented below.

**Lemma 3.** If $x$ be any member of the set $L_i - \{u_i'\}$ such that $(x, u_i') \notin E$ and $(x,y) \in E$ where $y \in L_{i+1} - \{u_{i+1}'\}$ then $(y, u_i') \in E$.

**Lemma 4.** If $x \in S_{i,(i-1)}'$, $y \in S_{i,(i-1)}' \cup S_{i,(i-1)}''$ and $(x,z) \in E$ where $z \in L_{i+1} - \{u_{i+1}'\}$ then $(y,z) \in E$.

**Proof.** Let $x$ be any member of the set $S_{i,(i-1)}'$ and $y$ be any member of the set $S_{i,(i-1)}' \cup S_{i,(i-1)}''$.

So in the trapezoid diagram $b_x < b_y$ as $(x, u_{i+1}') \notin E$. (3)

Again $(x,z) \in E$ where $z \in L_{i+1} - \{u_{i+1}'\}$. Therefore $b_z < a_x < b_x$. (4)

So from (1) and (2), we have $b_z < b_x < b_y$. This implies that $(y,z) \in E$.

**Lemma 5.** If $x \in S_{i,(i-1)}' \cup S_{i,(i-1)}''$ and $(y, u_{i+1}') \notin E$ where $y \in L_{i+1} - \{u_{i+1}'\}$ then $(x,y) \in E$.

**Proof.** Let $x$ be any member of the set $S_{i,(i-1)}' \cup S_{i,(i-1)}''$ then $(x, u_{i+1}') \in E$.

So, either $a_{u_{i+1}'} < b_x$ or $c_{u_{i+1}'} < d_x$ or both. (5)

Now $(y, u_{i+1}') \notin E$ where $y \in L_{i+1} - \{u_{i+1}'\}$. So in the trapezoid diagram, the trapezoid corresponding to the vertex $y$ will be scanned first than the trapezoid corresponding to the vertex $u_{i+1}'$ ( by the Algorithm TBFS [16]).

So, $b_y < a_{u_{i+1}'}$ and $d_y < c_{u_{i+1}'}$. (6)

Therefore from (1) and (2), we have $b_y < a_{u_{i+1}'} < b_x$ or $d_y < c_{u_{i+1}'} < d_x$. This implies that $(x,y) \in E$.

**Lemma 6.** If $(z,x) \notin E$ where $z \in D_i$, $x \in S_{i,(i-1)}^* - D_i$ then there exists at least one member $y \in L_{i+1}$ such that $(y,x) \in E$ for all $x \in S_{i,(i-1)}^* - D_i$.

**Lemma 7.** *If* $u^*_{i-1} \to u'_i \to u'_{i+1}$ *be a part of the main path (See Figure 5) and* $(x,y) \in E$ *but* $(y, u'_{i+1}) \notin E$ *where* $x \in S^*_{i,(i-1)}$, $y \in L_{i+1} - \{u'_{i+1}\}$ *then*
$u^*_{i-1} \to u^*_i (= u'_i) \to u_{i+1}(= u'_{i+1})$ *will be a part of the modified main path.*

**Lemma 8.** *If* $u^*_{i-1} \to u'_i \to u'_{i+1}$ *be a part of the main path and* $(z,x) \notin E$ *but* $(x,y) \in E$, $(y, u'_{i+1}) \in E$ *where* $z \in D_i$, $x \in S^*_{i,(i-1)} - D_i$ *and* $y \in P_{i+1} - \{u'_{i+1}\}$ *then* $u^*_{i-1} \to u^*_i (= u'_i) \to u_{i+1}$ *will be a part of the modified main path where* $b_{u_{i+1}} = max(b_i)$ *or* $d_{u_{i+1}} = max(d_i)$.

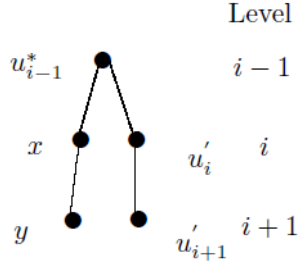

**Figure 5:** A part of the BFS tree $T^*(1)$.

**Lemma 9.** *If* $u^*_{i-1} \to u'_i \to u'_{i+1}$ *be a part of the main path and* $(x,y) \in E$, $(y,z) \in E$ *and* $(z, u'_{i+1}) \in E$ *where* $x \in D_i$, $y \in P_i - D_i - \{u'_i\}$ *and* $z \in P_{i+1} - \{u'_{i+1}\}$ *then* $u^*_{i-1} \to u^*_i \to u_{i+1}$ *will be a part of the modified main path where* $b_{u^*_i} = max(b^*_i)$ *or* $d_{u^*_i} = max(d^*_i)$ *and* $b_{u_{i+1}} = max\{b_z : z \in P_{i+1}$ *and* $(z, u^*_i) \in E\}$ *or* $d_{u_{i+1}} = max\{d_z : z \in P_{i+1}$ *and* $(z, u^*_i) \in E\}$.

**Lemma 10.** *If* $S_{1,0} = \phi$ *then* $u^*_0(= u'_0) \to u'_1 \to u'_2$ *can be taken as a part of the modified main path.*

## 5. The Algorithm
### 5.1. The modified main path

In Section 2, we construct a BFS tree $T^*(1)$ of the trapezoid graph $G$ and compute the main path. But it is obvious that $T^*(1)$ may or may not be a tree 3-spanner. So, for this purpose we modify the main path as well as the tree $T^*(1)$ with the help of the lemmas 7, 8 and 9. The modified tree is denoted by $T(1)$. the tree $T(1)$ is obtained from $T^*(1)$ by interchanging some or all edges of the main path of $T^*(1)$ with other edges of the graph $G$. Thus the main path of $T^*(1)$ has been changed and the changed main path is called the modified main path or the main path of $T(1)$. The modification can be done by the algorithm TR **3**SPT which is discussed in the next subsection.

### 5.2. The Algorithm

To find the tree 3-spanner on trapezoid graphs we first construct a BFS tree $T^*(1)$ with root as 1 and find the main path. Also we assume that $u_0^* = 1$ be the initial member of the modified main path as it is the root of the tree $T^*(1)$. Then we modify the BFS tree $T^*(1)$ to construct a tree 3-spanner which is denoted by $T(1)$. The main algorithm to find a tree 3-spanner of a trapezoid graph is presented below.

**Algorithm TR 3SPT**

**Input:** A trapezoid graph $G$ with the corner points $[a_i, b_i, c_i, d_i]$ of the trapezoid $i$ for all $i = 1, 2, \cdots, n$.

**Output:** Tree 3-spanner $T(1)$ of the trapezoid graph $G$.

**Step 1.** Construct a BFS tree $T^*(1)$ with root as 1 and let
$u_0' \to u_1' \to u_2' \to \cdots \to u_k'$ be the main path between 1 and $n$, where $1 = u_0'$ and $n = u_k'$.

**Step 2.** Compute the sets $L_i$ for $i = 1, 2, \cdots, h$.

**Step 3.** Mark all alternative shortest paths between $u_0'$ and the members of the set $L_h$.

**Step 4.** Compute the sets $P_i, F_i$ for $i = 1, 2, \cdots, h$.

**Step 5.** Let $u_0^* \to u_1' \to u_2'$ be a part of the *main path* where $u_0^* = u_0'$ and compute the sets $S_{1,0}$, $S_{1,0}'$, $S_{1,0}''$ and $S_{1,0}^*$.

**Step 6.** If $S_{1,0} = \phi$ or $S_{1,0} \neq \phi$ and $(x, y) \in E$, $(y, u_2') \notin E$ where
$x \in S_{1,0}^*, y \in P_2 - \{u_2'\}$, then $u_0^* \to u_1^* \to u_2$ will be the the part of the modified main path where $u_1^* = u_1'$ and $u_2 = u_2'$ (by Lemma 7, Lemma 10).

Else if $(z, x) \notin E$, $(x, y) \in E$ and $(y, u_2') \in E$ where $z \in D_i$, $x \in S_{1,0}^*$ and $y \in P_2 - \{u_2'\}$ then $u_0^* \to u_1^* \to u_2$ will be a part of the modified main path where $u_1^* = u_1'$ and $b_{u_2} = max(b_1)$ or $d_{u_2} = max(d_1)$
(by Lemma 8).

Else if $(x, y) \in E$, $(y, z) \in E$ and $(z, u_2') \in E$ where $x \in D_1$, $y \in P_1 - D_1 - \{u_1'\}$ and $z \in P_2 - \{u_2'\}$ then $u_0^* \to u_1^* \to u_2$ will be a part of the modified main path where $b_{u_1^*} = max(b_1^*)$ or $d_{u_1^*} = max(d_1^*)$ and

$$b_{u_2} = max\{b_z : z \in P_2 \text{ and } (z, u_1^*) \in E\} \text{ or}$$

$$d_{u_2} = max\{d_z : z \in P_2 \text{ and } (z, u_1^*) \in E\} \text{ (by Lemma 9).}$$

**Step 7.** Set $parent(x) = u_0^*$ where $x \in L_1 - \{u_1^*\}$ and $(x, u_1^*) \notin E, (x, u_2) \notin E$

and compute the set $C_{1,0} = \{x : x \in L_1 - \{u_1^*\} \text{ and } parent(x) = u_0^*\}$.

**Step 8.** Set $parent(y) = u_1^*$ where $y \in L_1 - \{u_1^*\} - C_{1,0}$, $(y, u_1^*) \in E$ and

$(y, x) \in E$ where $x \in C_{1,0}$ and compute the set

$C_{1,1} = \{x : x \in L_1 - \{u_1^*\} \text{ and } parent(x) = u_1^*\}$.

**Step 9.** Set $i = 2$ and if $i < h$ then go to next step, else go to Step 17.

**Step 10.** Let $u_{i-1}^* \to u_i' \to u_{i+1}'$ be a part of the *main path* where

$u_i' = u_i$ and $b_{u_{i+1}'} = max\{b_x : x \in P_{i+1} \text{ and } (x, u_i') \in E\}$ or

$d_{u_{i+1}'} = max\{d_x : x \in P_{i+1} \text{ and } (x, u_i') \in E\}$.

**Step 11.** Compute the sets $S_{i,(i-1)}$, $S'_{i,(i-1)}$, $S''_{i,(i-1)}$ and $S^*_{i,(i-1)}$.

**Step 12.** If $(x, y) \in E$, $(y, u_{i+1}') \notin E$ where $x \in S^*_{i,(i-1)}, y \in P_{i+1} - \{u_{i+1}'\}$,

then $u_{i-1}^* \to u_i^* \to u_{i+1}$ will be a part of the modified main path where

$u_i^* = u_i'$ and $u_{i+1} = u_{i+1}'$ (by Lemma 7).

Else if $(z, x) \notin E$, $(x, y) \in E$ and $(y, u_{i+1}') \in E$ where $z \in D_i$,

$x \in S^*_{i,(i-1)}$, $y \in P_{i+1} - \{u_{i+1}'\}$ then $u_{i-1}^* \to u_i^* \to u_{i+1}$ will be a part of the

modified main path where $u_i^* = u_i'$ and $b_{u_{i+1}} = max(b_i)$ or

$d_{u_{i+1}} = max(d_i)$ (by Lemma 8).

Else if $(z, x) \in E$, $(x, y) \in E$ and $(y, u_{i+1}') \in E$ where $z \in D_i$,

$x \in S^*_{i,(i-1)}$, $y \in P_{i+1} - \{u_{i+1}'\}$ then $u_{i-1}^* \to u_i^* \to u_{i+1}$ will be a part of the

modified main path where

$b_{u_i^*} = max(b_i^*)$ or $d_{u_i^*} = max(d_i^*)$ and

$b_{u_{i+1}} = max\{b_z : z \in P_{i+1} \text{ and } (z, u_i^*) \in E\}$ or

$d_{u_{i+1}} = max\{d_z : z \in P_{i+1} \text{ and } (z, u_i^*) \in E\}$ (by Lemma 9).

**Step 13.** If $(x, u_i^*) \in E$ where $x \in L_{i-1} - C_{(i-1),(i-2)} - C_{(i-1),(i-1)} - \{u_{i-1}^*\}$ then set

$parent(x) = u_i^*$ and compute the sets $C_{(i-1),(i)} = \{x : x \in L_{i-1} - \{u_{i-1}^*\} \text{ and}$

$parent(x) = u_i^*\}$.

Else set $parent(x) = u_{i-1}^*$ and compute the sets

Computation of a Tree 3-Spanner on Trapezoid Graphs

$C_{(i-1),(i-1)} = C_{(i-1),(i-1)} \cup \{x : x \in L_{i-1} - C_{(i-1),(i-2)} - C_{(i-1),(i-1)} - \{u^*_{i-1}\}$ and $parent(x) = u^*_{i-1}\}$.

**Step 14.** Set $parent(x) = u^*_{i-1}$ where $x \in L_i - \{u^*_i\}$ and $(x, u^*_i) \notin E, (x, u_{i+1}) \notin E$
and compute the sets $C_{i,(i-1)} = \{x : x \in L_i - \{u^*_i\}$ and $parent(x) = u^*_{i-1}\}$.

**Step 15.** Set $parent(y) = u^*_i$ where $y \in L_i - \{u^*_i\} - C_{i,(i-1)}$, $(y, u^*_i) \in E$ and
$(y, x) \in E$ where $x \in C_{i,(i-1)}$ and compute the sets
$C_{i,i} = \{y : y \in L_i - \{u^*_i\}$ and $parent(x) = u^*_i\}$.

**Step 16.** Set $i = i+1$.

**Step 17.** If $i = h$ then
if $(x, u^*_h) \in E$ and $(y, u^*_{h-1}) \in E$ where
$x \in L_{h-1} - C_{h-1,h-2} - C_{h-1,h-1} - \{u^*_{h-1}\}$ and $y \in L_h - \{u^*_h\}$ then set
$parent(x) = u^*_h$, $parent(y) = u^*_{h-1}$.
Else set $parent(x) = u^*_{h-1}$ and $parent(y) = u^*_h$.
Else go to Step 10.

**end TR3SPT.**

Using **Algorithm TR 3SPT** we get a tree, denoted by $T(1)$ which is shown in Figure 6. Next we are to show that the tree $T(1)$ is a tree 3-spanner.

It can be shown that the tree T(1) is a tree 3-spanner.

**Lemma 11.** *The tree $T(1)$ is a tree 3-spanner.*

Next we shall discuss about the time complexity of the **Algorithm TR3SPT** through following theorem.

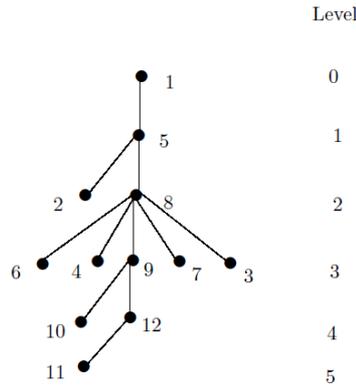

**Figure 6:** Tree 3-spanner $T(1)$ of the graph G of Figure 2.

**Theorem 2.** *The time complexity to find a tree 3-spanner on trapezoid graphs is $O(n^2)$, where $n$ is the number of vertices.*

**Proof.** A BFS tree $T^*(1)$ and the main path can be computed in $O(n)$ time, in Step 1. Step 2 can be computed in $O(n)$ time. Marking of all alternative shortest paths between $u_0^{'}$ and the members of the set $L_h$ can be computed in $O(n^2)$ time, in Step 3. The time complexity to compute the sets $P_i, F_i$ for $i = 1, 2, \cdots, h$, in Step 4, is $O(n)$. Step 5 can be completed in $O(n^2)$ time. The running time of Step 6 is $O(n^2)$. Step 7, can be finished in $O(n^2)$ time. Also the time complexity of the Step 8 is $O(n^2)$. The time complexity of the Step 9 is constant time. Step 10 can be completed in $O(n)$ time. In Step 11, the sets $S_{i,(i-1)}, S_{i,(i-1)}^{'}$, $S_{i,(i-1)}^{''}$ and $S_{i,(i-1)}^{*}$ can be computed in $O(n^2)$ time. Also Step 12 can be completed in $O(n^2)$ time. The time complexity of each step, Step 13, Step 14 and Step 15 is of $O(n^2)$. Step 16 can be run in constant time. The time complexity of Step 17 is $O(n^2)$. Hence, the over all time complexity of **Algorithm TR 3SPT** is $O(n^2)$.